\documentclass[slac_one]{revtex4}
\usepackage{graphicx}
\usepackage{fancyhdr}
\usepackage{xspace}
\newcommand{\dzero}{D\O~}
\newcommand{\ppbar}{\ensuremath{p\bar{p}}\xspace}
\newcommand{\qqbar}{\ensuremath{q\bar{q}}\xspace}
\newcommand{\ttbar}{\ensuremath{t\bar{t}}\xspace}
\newcommand{\mtop}{\ensuremath{m_{\rm top}}\xspace}
\newcommand{\ftop}{\ensuremath{f_{\rm top}}\xspace}
\newcommand{\psgn}{\ensuremath{P_{\rm sgn}}\xspace}
\newcommand{\pbkg}{\ensuremath{P_{\rm bkg}}\xspace}
\newcommand{\pevt}{\ensuremath{P_{\rm evt}}\xspace}
\newcommand{\etmiss}{\ensuremath{E \kern-0.6em\slash_{T}}\xspace}
\newcommand{\etmissxobs}{\ensuremath{E \kern-0.6em\slash_{x}^{obs}}\xspace}
\newcommand{\etmissyobs}{\ensuremath{E \kern-0.6em\slash_{y}^{obs}}\xspace}
\newcommand{\etmissx}{\ensuremath{E \kern-0.6em\slash_{x}}\xspace}
\newcommand{\etmissy}{\ensuremath{E \kern-0.6em\slash_{y}}\xspace}
\newcommand{\etmissxsmall}{\ensuremath{E \kern-0.4em\slash_{x}}\xspace}
\newcommand{\etmissysmall}{\ensuremath{E
    \kern-0.4em\slash_{y}}\xspace}
\begin{document}

\title{Measurements of the Top Quark Mass in the Dilepton Decay Channel \\
at the \dzero Experiment} %% Paper title goes here


\author{A. Grohsjean (for the \dzero collaboration)}
\affiliation{Ludwig-Maximilians-Universit\"at M\"unchen, 85748 Garching, Germany}

\begin{abstract}
We present the most recent measurements of the top quark mass in the
dilepton decay channel at the \dzero experiment using \ppbar collisions
with a center-of-mass energy of 1.96 TeV at the Tevatron collider. 
Two different methods have been used: the Neutrino Weighting 
and the Matrix Element method. The combined results yield a top mass of $174.4\pm3.8$~GeV.
\end{abstract}

\maketitle

\thispagestyle{fancy}

\section{INTRODUCTION} 
Since the discovery of the top quark in 1995 by the CDF and \dzero
experiments ~\cite{bib:CDFDISCOVERY,bib:D0DISCOVERY}, 
the mass of the top quark has been measured with ever 
higher precision. The mass is an important parameter within the
standard model (SM) and allows to infer the mass of the yet unobserved
Higgs boson. In addition, the top quark is the heaviest of the 
quarks known so far and with its Yukawa coupling close to one, it may play a
special role in electroweak symmetry breaking. 

At the Tevatron collider with a center-of-mass energy of 1.96 TeV, 
85~\% of the top quark pairs are produced in quark-antiquark annihilation,
while only 15~\% originate from gluon pair fusion. Since
the top quark decays to almost 100~\% in a charged $W$ boson and a
bottom quark, the decay channels of the top-antitop events are classified
according to the decays of the two $W$ bosons. The decay channel where each $W$
boson decays into a charged lepton and a neutrino is referred to as
dilepton channel. The topology of this channel is described by two 
isolated energetic charged leptons, significant missing transverse
energy, and two jets. 
The channel has the smallest branching ratio, but
also the smallest background contamination. The main backgrounds
to this final state are $Z$+jets and diboson events ($WW,WZ,ZZ$). 
Additional background comes from events in which a jet is
misidentified as an electron and the production of heavy hadrons 
that decay into leptons which pass the isolation requirements.
 
A systematic difference between top quark masses measured from
different decay channels could indicate contributions from new
processes beyond the SM.
The reconstruction of the top mass from dilepton events poses a 
particular challenge as the two neutrinos from the $W$ boson decays 
are undetected. At the \dzero experiment, two different methods are
used to measure the top mass: the Neutrino Weighting method
\cite{bib:LLNUWEIGHT08} and the
Matrix Element method \cite{bib:LLME08}. 
A brief introduction of both methods is given 
and their latest results are presented. 

\section{THE NEUTRINO WEIGHTING METHOD}

The Neutrino Weighting method is a template based method and has
already been used during the so-called Run~I period of the Tevatron accelerator.
For each event, the neutrino momenta are calculated assuming
a certain top mass and different neutrino pseudorapidities. 
A weight $w$ is then assigned according to the agreement of the calculated
sum of the neutrino momenta $\sum_{i=1}^2
p^{\nu_{i}}_x$, $\sum_{i=1}^2
p^{\nu_{i}}_y$, and the measured missing
transverse momentum components \etmissx,\etmissy in the event, given by
\begin{equation}
\label{eq:nuweight}
       w = \exp (\frac{-(\etmissx
	 - \sum_{i=1}^2 p^{\nu_{i}}_x)^2}{2\sigma_{\etmissxsmall}^2 }) 
       \exp (\frac{-(\etmissy
	 - \sum_{i=1}^2 p^{\nu_{i}}_y)^2}{2\sigma_{\etmissysmall}^2 })
       \,,
\end{equation}
where $\sigma_{\etmissxsmall},\sigma_{\etmissysmall}$ denote the
resolution of the missing energy measurement.
%\frac{1}{n_{{\rm sol}}} \sum^{n_{{\rm sol}}} 
This process is repeated many times varying the jet and lepton
energies within their experimental resolution. Next, 
signal probability distribution functions (PDF) are built  
as a function of the top mass,
the mean, and the RMS of the weight distributions using Monte Carlo
(MC) simulated signal events with different top masses. To reduce the bias
from background contamination, a background PDF as a function of the mean
and the RMS of the weight distributions is calculated accordingly
using simulated $Z$+jets and diboson events. 
Finally, a likelihood is used to measure the top mass, where the
likelihood is a product of three terms. The first term accounts for the agreement
of the expected number of signal and background MC events to the one in the data sample,
the second for the agreement of the background events with the
prediction, and the third for the agreement of the data with the
signal and background PDF shapes. 
The method is calibrated using ensemble tests with Monte Carlo
simulated events, where the size of the ensemble is equal to the
size of the selected data sample.   

For the measurement, events in the dilepton final state are selected requiring two
charged leptons (electron or muon), or
alternatively one lepton, one charged particle track and at least one $b$ tagged
jet. The data set analyzed corresponds
to an integrated luminosity of about 1~fb$^{-1}$ collected between April
2002 and February 2006 with the \dzero detector. A kinematic selection is applied to
reduce the contamination from background events and to achieve a good
agreement between simulated and measured events. The top quark mass 
is extracted from 82 candidate events to be 
\begin{equation}
\mtop=176.0\pm5.3({\rm stat})\pm2.0({\rm syst}) {\rm GeV} .
\end{equation}
The main systematic uncertainties on this measurement come from
the energy scale of the jets, the modeling of the simulated signal
events using different MC generators, and the fragmentation of the $b$ jets. 

\section{THE MATRIX ELEMENT METHOD}
The Matrix Element method was developed by the \dzero collaboration
during the Run~I period of the Tevatron to
extract the top mass with high precision from a low statistics
sample. 
Still today the Matrix Element method yields the 
single most precise measurement of the top quark mass in the semileptonic
decay channel. 

The likelihood for a measured event
$x$ to be produced via the signal process $\ppbar \rightarrow \ttbar
\rightarrow y$ under the assumption of a certain top mass is given by
\begin{equation}
\label{eq:psgn}
\psgn(x;\mtop) = \frac{1}{\sigma_{\rm obs}(\mtop)}\cdot \int_{y} 
~{\rm d}\epsilon_1 {\rm d}\epsilon_2
f(\epsilon_1)f(\epsilon_2)
 ~\frac{(2 \pi)^{4} \left|M(\qqbar \rightarrow \ttbar \rightarrow y )\right|^{2}}
       {\epsilon_1\epsilon_2 s}
 ~      {\rm d}\Phi_{6}
\cdot W(x,y) \ , 
\end{equation}
where $\epsilon_1$, $\epsilon_2$ denote the energy fraction of the
incoming quarks from the protons and antiprotons, $f$ the parton distribution function
of the proton, $s$ the center-of-mass energy squared, 
$M(\qqbar \rightarrow \ttbar \rightarrow y)$ the leading-order matrix
element \cite{bib:MAHLON} and ${\rm d}\Phi_{6}$ an element of the 6-body phase space.  
The finite detector resolution is taken into account using a transfer
function $W(x,y)$ that describes the probability to reconstruct a
partonic final state $y$ as $x$ in the detector. The signal
probability is normalized with the observable cross section
$\sigma_{\rm obs}$.

In a similar way, for each event the probability to arise from the main
background source (Z+2jets) is calculated to reduce the bias
coming from background events. Both probabilities are then combined to
an event probability  
 \begin{equation}
  \pevt(x;\mtop) = \ftop\cdot\psgn(x;\mtop)+(1-\ftop)\cdot\pbkg(x) \, ,
\end{equation}
where \ftop denotes the signal fraction in the sample.
The top quark mass is finally extracted from a likelihood fit of the
product of the event by event probabilities.  
To calibrate the method and correct for any bias, Monte
Carlo simulated events are used to perform ensemble tests. 

While only events with exactly one electron and exactly one muon in the final state
are taken into account so far, the measurement makes use of the full
Run~II data set recorded between April 2002 and May 2008. This
corresponds to an integrated luminosity of 2.8~fb$^{-1}$. To reduce the
fraction of background events and achieve a good agreement between
simulated and measured events, a kinematic selection is applied,
leaving 107 selected data events.
 
The top quark mass is measured  to be 
\begin{equation}
\mtop=172.9\pm3.6({\rm stat})\pm2.3({\rm syst}) {\rm GeV} .
\end{equation}
The dominant sources of systematic uncertainties are  
jet uncertainties, such as their energy scale and resolution.
With an statistical uncertainty of $3.6$~GeV, this measurement has the
smallest uncertainty of all top mass measurements performed in the
dilepton channel at the \dzero experiment so far.

\section{CONCLUSION}
\begin{figure*}[h]
\centering
\includegraphics[width=60mm]{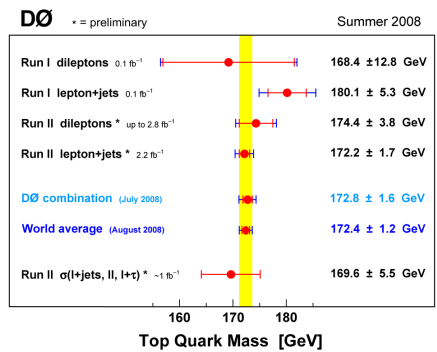}
\caption{Combination of the Run~I and Run~II measurements of the top
  quark mass at the \dzero experiment in the lepton+jets and dilepton decay channel, as
  well as the current world average. } \label{fig:topmass0808}
\end{figure*}

The top quark mass has been measured at the \dzero experiment with two
different methods in the dilepton final state: the Neutrino Weighting
and the Matrix Element method. Uncertainties from the jet energy scale
dominate the systematic uncertainties in both measurements.  
The Best Linear Unbiased Estimate (BLUE) method is used to combine
both measurements \cite{bib:LLME08,bib:COMBI08}. All uncertainties are
taken to be fully correlated except for the statistical and
fit uncertainties from the calibration curves. 
To combine the results of two orthogonal samples, the result from the Neutrino
Weighting method in the channel with exactly one electron and
exactly one muon is excluded and the preliminary combined result of the two
methods yields a top quark mass of  
\begin{equation}
\mtop=174.4\pm3.2({\rm stat})\pm2.1({\rm syst}) {\rm GeV} .
\end{equation}
The current measurements and uncertainties on the top quark mass at
the \dzero experiment, as well as the world average are shown in
Figure \ref{fig:topmass0808} \cite{bib:COMBI08}.
In a future update of the Matrix Element measurement using a larger
data set, statistical uncertainties in the dilepton final state will be
reduced significantly. Thus, measurements in the dilepton channel will
become more and more important determining the world average of the
top quark mass.

%\begin{acknowledgments}n
%The authors wish to thank JACoW for their guidance in preparing
%this template.
%\end{acknowledgments}

\end{document}